\documentclass[prd,showpacs,showkeys]{revtex4}
\usepackage[latin9]{inputenc}
\usepackage{amsmath}
\usepackage{amssymb}
\usepackage{graphicx}

\usepackage{graphics}
\usepackage{eurosym}
\usepackage{amsfonts}
\usepackage{verbatim}

\begin{document}

\title{Particle Acceleration by Static Black Holes in a Model of $f(R)$
Gravity}

\author{M. Halilsoy}
\email{mustafa.halilsoy@emu.edu.tr}

\author{A. Ovgun}
\email{ali.ovgun@emu.edu.tr}

\affiliation{Physics Department, Eastern Mediterranean University, Famagusta,
Northern Cyprus, Mersin 10, Turkey.}

\date{\today }

\begin{abstract}
Particle collisions are considered within the context of $f(R)$ gravity
described by $f(R)=R+2\alpha\sqrt{R}$, where $R$ stands for the
Ricci scalar and $\alpha$ is a non-zero constant. The center of mass
(CM) energy of head-on colliding particles moving in opposite radial
directions near the naked singularity/horizon are considered. Collision
of particles in the same direction near the event horizon yields finite
energy while the energy of oppositely moving particles grows unbounded.
Addition of a cosmological constant does not change the feature. Collision
of a massless outgoing photon with an infalling particle and collision
of two oppositely moving photons following null-geodesics are also
taken into account. 

\end{abstract}

\maketitle

\section{Introduction}

In particle accelerators, physicists routinely accelerate elementary
particles and bring them to collision. Banados, Silk and West (BSW)
\cite{BSW} have proposed a scenario, where black holes may act as
particle accelerators and first showed that the collision of geodesic
particles in the vicinity of a black hole horizon yields a total unbounded
centre of mass (CM) energy. This amounts to a natural collision process
similar to the artificially tested process in the high-energy laboratory
at CERN. The difference is that the latter is under strict human control
albeit a bit too expensive process whereas the former one is free
of charge, occurring in cosmos frequently as an ordinary event. Not
only do the black holes create a similar BSW effect, but also naked
singularities as well as the throat regions of wormholes \cite{bambi1}.
Rotating black holes and wormholes act more efficiently in comparison
with static ones to yield a high CM energy \cite{bobir,pradhan,pour,sadeghi,josji,joshi2,pavlov,stuchlik,gao,liu}.
Another aspect of the BSW effect is that it occurs irrespective of
the dimensionality of spacetime or the nature of the underlying theory.
That is, even in lower/higher dimensions of 3+1- spacetime we can
have an accelerator effect\cite{schh,ayesha,josji3,wei,adami,harada,malafarina,west,anton,sheoran,piran,sotiriou,lake,nunes,grib,grib2,chenn,ghoshh,debnath,jing,dingliu,horava,ali1,Berti,elly}.
Collision must take place near the horizon of the formed black hole
or naked singularity so that the particles get boost from the unlimited
attraction/repulsion \cite{BSW,tomira,jamil,sharif}. 

The main aim of the paper is to study the BSW process which
is useful to detect relic cold dark matter particles which are located
around the black holes. Massive dark matter particles collide each
others and the center of mass energy may reach arbitrarily high energies
\cite{BSW}. In our earlier study \cite{ali1} we showed that modification
of the Einstein gravity, Horava-Lifshitz gravity, does not always
lead to infinite CM energy. Since so far the BSW effect is
found, first time in this paper we investigate the possibility of
BSW effect in the modified Einstein theory known as the $f(R)$ gravity
with static, diagonal metrics \cite{sebastiani}. In this theory the
Einstein-Hilbert action characterized by the Ricci scalar $(R$) is
extended to cover an arbitrary function of $R$. Theoretically such
a concept has an infinite number of extensions which are to be severely
restricted by experimental tests. Naturally any higher power of $R$
hosts higher order derivatives of the metric and expectedly obtaining
exact solutions is not an easy task at all. The solution for $f(R)$
gravity that we shall consider in this study is a static one with
$f(R)=R+2\alpha\sqrt{R-4\Lambda}-2\Lambda$, in which the constant
$\alpha\neq0$, so that our model of $f(R)$ has no vacuum Einstein
limit. By comparison with the Schwarzchild - de Sitter line element
the second integration constant $\Lambda$ can be interpreted as a
cosmological constant. Since it is expected that most rotating
black holes would have faced the BSW effect, in fact modification
of the Einstein theory changes the structure of the black hole and
the BSW effect in some cases occur and dark matter particles is accelerated.
We concentrate here on our main new result, namely computation of
the limiting energy for $f(R)$ black holes. In this paper we also address
the issue of CM of high energy collisions in the absence of an event
horizon and near the naked singularity.

We consider first $\alpha<0$ case which represents a black hole in
which collision of two infalling particles takes place near the event
horizon. Oppositely moving particles near the event horizon does yield
BSW effect, however, the physical situation prohibits the existence
of outgoing particles from the event horizon \cite{Berti,elly}. For
that reason we base our argument on some physical processes that involve
decay/ disintegration of particles outside the horizon. Once such
a process is assumed valid there will be outgoing as well as infalling
particles in the vicinity of a black hole. As a result the existence
of outgoing particles/photons will naturally invite the process of
collision with different infalling particles/photons. Next, we consider
the case $\alpha>0$, as a naked singularity at $r=0$ and the collision
of two oppositely moving particles near $r\thickapprox0$. It is found
that due to the physical constraints, such as real momenta no unbounded
CM energy arises from collisions in the vicinity of naked singularity.
We note that the outgoing particle may be attributed due to the repulsive
effect of the naked singularity which reverses/rebounds the particles
and photons from $r\thickapprox0$ . We cite as an example the case
of negative mass Schwarzschild metric which gives rise to repulsive
gravitational effect. Particles/photons turning outward can naturally
make geodesic collisions with the incoming particles.

The paper is organized as follows: Section II summarizes static spherically
symmetric black holes and naked singularities in a model of f(R) gravity.
Collision of particles, between outgoing and infalling particles near
the naked singularities and event horizons are analysed in Section
III. Section IV considers collisions involving photons, both outgoing
and infalling. We complete the paper with our conclusion in Section
V.

\section{Specific Black Hole/ Naked singularity}

The action of general, sourceless $f(R)$ gravity theories in four
dimension is 
\begin{equation}
S=\frac{1}{16\pi G}\int\text{d}^{4}x\sqrt{-g}\,f(R),\label{1}
\end{equation}
in which $f(R)$ is the function of the scalar curvature $R$ ,and
$g$ stands for the determinant of the metric tensor \cite{sebastiani}.

The spherically symmetric line element is given by

\begin{equation}
ds^{2}=-Adt^{2}+\frac{1}{A}dr^{2}+r^{2}(d\theta^{2}+\sin^{2}\theta d\phi^{2}),
\end{equation}
in which the function of $f(R)$ 
\begin{equation}
f(R)=R+2\alpha\sqrt{R},
\end{equation}
with the constant $\alpha\neq0,$ yields the solution

\begin{equation}
A(r)=\frac{1}{2}+\frac{1}{3\alpha r},
\end{equation}
with the Ricci scalar

\begin{equation}
R=\frac{1}{r^{2}}.
\end{equation}
Inclusion of a cosmological constant $\Lambda$ yields the function
$f(R)$ as follows \cite{amirabi} 
\begin{equation}
f(R)=R+2\alpha\sqrt{R-4\Lambda}-2\Lambda,
\end{equation}
and the corresponding metric function of $A(r)$ is 
\begin{equation}
A(r)=\frac{1}{2}+\frac{1}{3\alpha r}-\frac{\Lambda}{3}r^{2},
\end{equation}
with the scalar curvature 
\begin{equation}
R=\frac{1}{r^{2}}+4\Lambda.
\end{equation}
Note that the fact that $\alpha\neq0$ is already revealed by the
metric function $A(r)$. In the sequel for both cases, $\Lambda=0$
and $\Lambda\neq0$ we shall investigate the possibility of BSW effect.
Lastly, for the case of $\alpha>0$, ($\Lambda=0$ ) which corresponds
to a naked singular solution at $r=0$ we shall search for the collider
effect. For the case of naked singularity, the metric function \ is
calculated (let $\Lambda=0$) as follows

\begin{equation}
A=\frac{1}{2}+\frac{1}{3\alpha r}.
\end{equation}

Obviously in the two parametric solution employed, $\Lambda$ is a
dispensable parameter whereas $\alpha$ not. That is, our choice of
$f(R)$ gravity lacks the Einstein's general relativity limit. With
deliberation we \ have made such a choice to see the significance
of the BSW effect in a $f(R)$ model that is not connected with the
general relativity. This is precisely the case with $\alpha\neq0$.

\section{Particle Collision near $f(R)$ Black Hole and Naked Singularity}

\begin{figure}[h]
\includegraphics[width=0.3\textwidth]{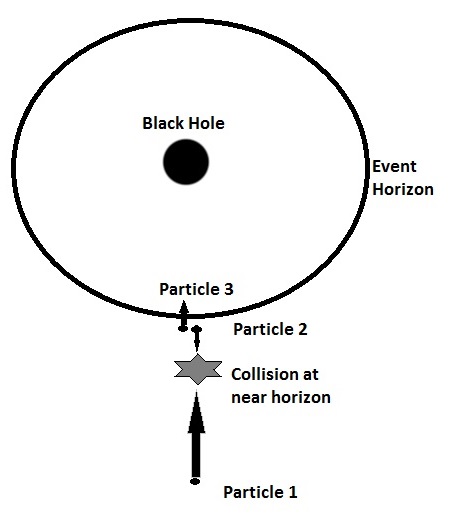} \caption{The schematic figure of particle collision for which the CM energy
can be very large (Particles 1 and 3 are infalling while particle
2 is outgoing) }
\label{bswplot} 
\end{figure}

We wish to check first the role of event horizon when particles collide
in case our metric is static/diagonal in $f(R)$ gravity. For different
cases we investigate the CM energy for the collision, \ 4-d velocity
components of the colliding particles in the background of the 4-d
$f(R)$ black holes by taking the radial motion on equatorial plane
($\theta=\frac{\pi}{2}$) (Fig. \ref{bswplot}).

Our Lagrangian is chosen by

\begin{equation}
L=\frac{1}{2}\left(-A\dot{t}^{2}+\frac{1}{A}\dot{r}^{2}+r^{2}\dot{\varphi}^{2}\right),
\end{equation}
in which a dot implies derivative with respect to proper time. The
velocities follow as 
\begin{equation}
u^{t}=\dot{t}=\frac{E}{A},
\end{equation}
and 
\begin{equation}
u^{\varphi}=\dot{\varphi}=\frac{L}{r^{2}},
\end{equation}
where E and L are the energy and angular momentum constants, respectively.
By using the normalization condition ($u.u=-1$), it is found that
the radial velocity is 
\begin{equation}
u^{r}=\dot{r}=\pm\sqrt{E^{2}-A\left(1+\frac{L^{2}}{r^{2}}\right)},
\end{equation}
and clearly we are interested in time-like geodesics. We proceed now
to present the CM energy of two particles with four-velocities $u_{1}^{\mu}$
and $u_{2}^{\mu}$. We assume that both have rest mass $m_{0}=1$.
The CM energy is given by, 
\begin{equation}
E_{cm}=\sqrt{2}\sqrt{(1-g_{\mu\nu}u_{1}^{\mu}u_{2}^{\nu})},\label{8}
\end{equation}
so that it can be expressed as 
\begin{equation}
\frac{E_{cm}^{2}}{2}=1+\frac{E_{1}E_{2}}{A}-\kappa\frac{|L_{1}||L_{2}|}{r^{2}}-\kappa\frac{1}{A}\sqrt{E_{1}^{2}-A\left(1+\frac{L_{1}^{2}}{r^{2}}\right)}\sqrt{E_{2}^{2}-A\left(1+\frac{L_{2}^{2}}{r^{2}}\right)},
\end{equation}
where $\kappa=\pm1$ correspond to particles moving in the same direction
$(\kappa=+1)$ or opposite direction $(\kappa=-1)$, respectively.
Furthermore $E_{1}$ and $E_{2}$/ $L_{1}$ and $L_{2}$ are defined
as the energy/ angular momentum constants corresponding to each particle.
Upon taking the lowest order terms in the vicinity of the horizon,
since $A\approx0$, we can make the expansion 
\begin{equation}
\sqrt{\left[E^{2}-A\left(1+\frac{L^{2}}{r^{2}}\right)\right]}\cong E\left[1-\frac{A}{2E^{2}}\left(1+\frac{L^{2}}{r^{2}}\right)+...\right],
\end{equation}
so that the CM energy of two particles is obtained as 
\begin{equation}
\frac{E_{cm}^{2}}{2}\cong1+(1-\kappa)\frac{E_{1}E_{2}}{A}-\kappa\frac{|L_{1}||L_{2}|}{r^{2}}+\frac{\kappa}{2}\left[\frac{E_{2}}{E_{1}}\left(1+\frac{L_{1}^{2}}{r^{2}}\right)+\frac{E_{1}}{E_{2}}\left(1+\frac{L_{2}^{2}}{r^{2}}\right)\right].\label{cm}
\end{equation}
We investigate the BSW effect whether occurs or not for $A(r)\rightarrow0$
\ whenever there is a horizon. In the case of the collision of ingoing/ingoing
or outgoing/outgoing particles (i.e. motion in same directions) $\kappa=+1$,
it reduces to 
\begin{equation}
\frac{E_{cm}^{2}}{2}\cong1-\frac{|L_{1}||L_{2}|}{r^{2}}+\frac{1}{2}\left[\frac{E_{2}}{E_{1}}\left(1+\frac{L_{1}^{2}}{r^{2}}\right)+\frac{E_{1}}{E_{2}}\left(1+\frac{L_{2}^{2}}{r^{2}}\right)\right].\label{eq:18}
\end{equation}

For the case of the $f(R)$ black hole without cosmological constant
where $\alpha<0$, horizon is at $r_{h}=\frac{2}{3|\alpha|},$\ and
$A$ goes to zero at horizon of $f(R)$ black hole, however Eq.(\ref{eq:18})
has not any term of A so that it does not diverge and there is no
BSW effect. On the other hand, when $\kappa=-1$, i.e when the particles
move in the opposite directions, we observe that the second term in
Eq. 17 becomes unbounded. We stress once more that if the particles
are moving both inward, i.e. $\kappa=+1$, there is no BSW effect
in the $E_{cm}^{2}$, in accordance with Eq. (\ref{eq:18}).

The occurrence of outgoing particles is crucial for a diverging $E_{cm}^{2}$.
Such an outgoing particle may be attributed to a decay/disintegration
process in the vicinity of the horizon. While one of the particle
falls into the hole its pair moves outward to collide with an infalling
particle.

\subsection{Particle Collision near the $f(R)$ Black Holes with a Cosmological
Constant}

The second case of interest is for the chosen $f(R)$ black hole model
with a cosmological constant in which the metric function A is 
\begin{equation}
A=\frac{-\Lambda r^{2}}{3}+\frac{1}{2}+\frac{1}{3\alpha r},\label{metricfunc}
\end{equation}
where the event horizon is located at 
\begin{equation}
r_{h}=\frac{\Xi}{2\alpha\Lambda}+\frac{\alpha}{\Xi},\label{hh}
\end{equation}
for 
\begin{equation}
\Xi=\left(4+2\alpha^{2}\Lambda^{2}\sqrt{\frac{-2\alpha^{2}-4\Lambda}{\Lambda}}\right)^{\frac{1}{3}}.
\end{equation}
At this point we must add that we are not interested in the other
roots of $A(r)=0$ that specify the inner horizon. It is observed
that for real $r_{h}$ we must have $\frac{-2\alpha^{2}}{\Lambda}-4>0$,
which restricts the cosmological constant to the case of $\Lambda<0$.

As in the case of $\kappa=+1$ above the CM energy $E_{c.m.}^{2}$is
finite. Collision of an infalling and outgoing particle $\kappa=-1$
, however, does yield a BSW effect.

\subsection{Particle Collision near the Naked Singularity }

There is a naked singularity for our $f(R)$ model at the location
of $r=0$ , with $\alpha>0$,where the metric function \ is given
by (let $\Lambda=0$)

\begin{equation}
A=\frac{1}{2}+\frac{1}{3\alpha r}.
\end{equation}
As it is calculated above the collision of two particles generally
is 
\begin{equation}
\frac{E_{cm}^{2}}{2}=1+\frac{E_{1}E_{2}}{A}-\kappa\frac{|L_{1}||L_{2}|}{r^{2}}-\kappa\frac{1}{A}\sqrt{E_{1}^{2}-A\left(1+\frac{L_{1}^{2}}{r^{2}}\right)}\sqrt{E_{2}^{2}-A\left(1+\frac{L_{2}^{2}}{r^{2}}\right)}.
\end{equation}

Let us note that each term under the square root must be positive.
Such a constraint restricts the range of r to stay away from the naked
singularity. Once r is finite the overall CM energy also must be finite
and therefore we observe no diverging result from the presence of
the naked singularity. Choosing the pure radial motions. (i.e.$L_{1}=L_{2}=0$)
also does not change the feature of the problem.

\section{Collisional Processes with Photons}

\subsection{In Naked Singular Spacetime}

The outgoing massless photon presumably reflected from the naked singularity
can naturally scatter an infalling particle or vice versa. This phenomenon
is analogous to a \ Compton- like scattering process which was originally
introduced for a photon and an electron.\ The null-geodesics for
a photon satisfies 
\begin{equation}
\frac{dt}{d\lambda}=\dot{t}=\frac{E_{\gamma}}{A},
\end{equation}
and 
\begin{equation}
\frac{d\varphi}{d\lambda}=\dot{\varphi}=\frac{L_{\gamma}}{r^{2}},
\end{equation}
\begin{equation}
\dot{r}=\pm\sqrt{\left[E_{\gamma}^{2}-\frac{AL^{2}}{r^{2}}\right]},
\end{equation}
where $\lambda$ and $E_{\gamma}$are \ an affine parameter and the
photon energy, respectively. Defining $E_{\gamma}=\hslash\omega_{0}$,
where $\omega_{0}$ is the frequency (with the choice $\hslash=1)$
we can parametrize the energy of the photon \ by $\omega_{0}$ alone.
The center-of-mass energy of an outgoing photon and the infalling
particle can be taken now as

\begin{equation}
E_{cm}^{2}=-(p^{\mu}+k^{\mu})^{2},
\end{equation}
in which $p^{\mu}$ and $k^{\mu}$ refer to the particle and photon,
4- momenta, respectively. It is needless to state that for a photon
we have $k^{2}=0$. This amounts to (for $\theta=\pi/2$) 
\begin{equation}
E_{cm}^{2}=m^{2}-2mg_{\mu\nu}p^{\mu}k^{\nu}.
\end{equation}

Since we have for the particle 
\begin{equation}
p^{\mu}=m\left(\frac{E_{1}}{A},\sqrt{E_{1}^{2}-A\left(1+\frac{L_{1}^{2}}{r^{2}}\right)},0,\frac{L_{1}}{r^{2}}\right),
\end{equation}
and for the photon 
\begin{equation}
k^{\mu}=\left(\frac{E_{\gamma}}{A},\sqrt{E_{\gamma}^{2}-\frac{AL_{\gamma}^{2}}{r^{2}}},0,\frac{L_{\gamma}}{r^{2}}\right),
\end{equation}
one obtains 
\begin{equation}
E_{cm}^{2}=m^{2}+\frac{2mE_{\gamma}E_{1}}{A}-\frac{2\kappa m|L_{1}||L_{\gamma}|}{r^{2}}-\frac{2m\kappa}{A}\sqrt{E_{\gamma}^{2}-\frac{AL_{\gamma}^{2}}{r^{2}}}\sqrt{E_{1}^{2}-A\left(1+\frac{L_{1}^{2}}{r^{2}}\right)}.
\end{equation}
which amounts to a collision process that must occur away from the
singularity $r=0$, i.e. $E_{cm}^{2}$ remains finite.

\subsection{Photon - Photon Collision Near the Naked Singularity}

Let us consider the problem of collision between two photons in the
naked singular spacetime. The photons follow null geodesics in opposite
directions and make head-on collision. In quantum electrodynamics
colliding energetic photons can transmute into particles. Since our
analysis here is entirely classical we shall refer only to the CM
energy of the yield without further specification. The CM energy of
the product satisfies

\begin{equation}
E_{cm}^{2}=-(k_{1}^{\mu}+k_{2}^{\mu})^{2}=-2g_{\mu\nu}k_{1}^{\mu}k_{2}^{\nu},\label{50}
\end{equation}
where $k_{1}$ and $k_{2}$ correspond to the 4- momenta of respective
photons. From the null- geodesic analysis in the $\theta=\frac{\pi}{2}$
plane, we have

\begin{equation}
k_{1}^{\mu}=\{\frac{E_{1}}{A},\sqrt{E_{1}^{2}-\frac{AL_{1}^{2}}{r^{2}}},0,\frac{L_{1}}{r^{2}}\},
\end{equation}

\begin{equation}
k_{2}^{\mu}=\{\frac{E_{2}}{A},\sqrt{E_{2}^{2}-\frac{AL_{2}^{2}}{r^{2}}},0,\frac{L_{2}}{r^{2}}\},
\end{equation}
where $E_{1}$ and $E_{2}$ are the corresponding energies of different
photons. Upon substitution into (\ref{50}) we obtain

\begin{equation}
\frac{1}{2}E_{cm}^{2}=\frac{E_{1}E_{2}}{A}-\frac{\kappa}{A}\sqrt{E_{1}^{2}-\frac{AL_{1}^{2}}{r^{2}}}\sqrt{E_{2}^{2}-\frac{AL_{2}^{2}}{r^{2}}}-\frac{|L_{1}||L_{\gamma}|}{r^{2}},
\end{equation}
in which we have to insert $\kappa=\pm1$ to specify the parallel/anti-parallel
propagation of the photons.

1) For $\kappa=+1$ , which implies two parallel photons moving at
the speed of light naturally don't scatter, so we observe no noticeable
effect.

2) For $\kappa=-1$ , however, the photons are moving in opposite
directions and inevitably they collide. In classical background each
naturally follows a null geodesics. Their corresponding CM energy
from our foregoing analysis that $r_{i}^{2}>\frac{AL_{i}^{2}}{E_{i}^{2}}$
for each $i=1,2$ remains also finite.

Let us comment that this is a collision of test photons on a given
geometry without backreaction effect. On the other hand, exact collision
of electromagnetic shock plane waves in Einstein's gravity, as a highly
non- linear process \cite{Bell} \cite{halilsoyy} is entirely different.
As a result of mutual focusing, the latter develops null- singularities
after the collision process. Our conclusion is that, at the test level,
collision of two oppositely moving photons in a naked singular spacetime
yields no observable effect.

\subsection{Photon - Photon Collision Near the Horizon}

From the analysis in part B above the CM energy of two photons is
adapted as

\begin{equation}
\frac{1}{2}E_{cm}^{2}=\frac{E_{1}E_{2}}{A}-\frac{\kappa}{A}\sqrt{E_{1}^{2}-\frac{AL_{1}^{2}}{r^{2}}}\sqrt{E_{2}^{2}-\frac{AL_{2}^{2}}{r^{2}}}-\frac{|L_{1}||L_{\gamma}|}{r^{2}},
\end{equation}
with the supplement that now we search the case for the limit $A\rightarrow0$,
instead of $A\rightarrow\infty$, since $r=r_{h}=finite.$ We obtain
\[
\frac{1}{2}E_{cm}^{2}\eqsim\frac{E_{1}E_{2}}{A}(1-\kappa)+(finite\text{ }terms),
\]
which suggests that for $\kappa=-1$, i.e. for oppositely moving photons,
it yields an unbounded CM energy. For parallel photons, both ingoing,
naturally $\kappa=+1$ and there is no observable effect. 

How can an outgoing photon from the near-horizon region can be justified
since the Hawking photons remains too weak to cope/scatter with an
infalling one?. An outgoing photon can be created in an explosion/
decay process by a physical particle before it falls into the horizon.
Such an assuption yields an outgoing photon and naturally it has the
chance to collide with an opposite photon and give rise to an unbounded
energy. This is exactly what happens for two colliding oppositely
moving electromagnetic plane waves to focus each other and create
a null singularity\cite{Bell,halilsoyy}.

\section{Conclusion}

Collision of particles near black hole horizons in Einstein's general
relativity, i.e. the BSW effect, has been considered in details during
recent years. Oppositely moving particle collisions near static black
holes were also considered in \cite{pavlov,prd}. Besides static,
charged and rotating black holes were also investigated.\ In particular,
rotational effects were shown from the original Penrose process long
ago \cite{penrose1}. 

This has a significant role in the extraction of energy from the black
holes. In this paper, we investigated the idea of BSW to the modified
theory known as $f(R)$ gravity. In particular we concentrated on
$f(R)=R+2\alpha\sqrt{R-4\Lambda}-2\Lambda$, which arises as an exact,
source- free spherically symmetric solution that the external energy-
momentum tensor vanishes, but the curvature makes its own source.
We can easily set $\Lambda=0$, however $\alpha\neq0$ is an essential
parameter of the model so that our model does not have the general
relativity limit of $f(R)=R$. For $\alpha<0$ we have the black hole
while for $\alpha>0$ we obtain a naked singularity at $r=0$. 

In case of black hole we show the existence of BSW effect provided
that outgoing particles from some physical process is taken for granted.
Collision of an infalling and outgoing particle $\kappa=-1$ near
the horizon of the$f(R)$ black holes with/ without a cosmological
constant, however, does yield a BSW effect. Near a naked singularity,
however, we observe no efficient collision to increase the CM energy
unbounded. For oppositely moving particles a similar result can also
be obtained for a Compton- like process between a photon and a particle
provided that they move in opposite directions. Collision of two oppositely
moving photons near the naked singularity also yields no diverging
CM energy. On the other hand, for oppositely moving photons,
it yields an unbounded CM energy. Therefore it is clear that CM energy
depends on the direction of the particles with the parameter of $\kappa$
and the collision of the oppositely moving particles must be near
the horizon of the black hole.

The CM energy distribution of relic cold dark matter particles
colliding in lower/higher dimensions will be discussed in a future
publication with comparing the observational datas that give the possible
excess of gamma rays observed in Fermi data at WIMP-scale energies
\cite{su}. Moreover, it is our belief that seeking an alternative
model of gravity, which can lead to BSW effect will be useful in the
searching of the dark matter. On this purpose we will investigate
the BSW process for black holes/ strings or wormholes to look at the
CM energy of the colliding neutral/charged particles. This is going
to be our next problem in the near future.

\end{document}